# Estimation of economic losses due to milk fever and efficiency gains if prevented: evidence from Haryana, India

A. G. Adeeth Cariappa*[1], B. S. Chandel[1], Gopal Sankhala[2], Veena Mani[3], Sendhil R[4], Anil Kumar Dixit[1] and B. S. Meena[2]

**Abstract**

Calcium (Ca) requirement increases tenfold upon parturition in dairy cows & buffaloes and its deficiency leads to a condition called milk fever (MF). Estimation of losses is necessary to understand the depth of the problem and design preventive measures. How much is the economic loss due to MF? What will be the efficiency gain if MF is prevented at the advent of a technology? We answer these questions using survey data and official statistics employing economic surplus model. MF incidence in sample buffaloes and cows was 19% and 28%, respectively. Total economic losses were calculated as a sum total of losses from milk production, mortality of animals and treatment costs. Yearly economic loss due to MF was estimated to be ₹ 1000 crores (US$ 137 million) in Haryana. Value of milk lost had the highest share in total economic losses (58%), followed by losses due to mortality (29%) and treatment costs (13%). Despite lower MF incidence, losses were higher in buffaloes due to higher milk prices and market value of animals. The efficiency gain accruing to producers if MF is prevented, resulting from increased milk production at decreased costs was estimated at ₹ 10990 crores (US$ 1.5 billion). As the potential gain if prevented is around 10 times the economic losses, this study calls for the use of preventive technology against MF.

# Introduction

Upon parturition or calving, due to increased milk and colostrum production in dairy cows and buffaloes, calcium (Ca) requirement will be 10 times more than the dry period (Patel et al. 2011). If this requirement is not fulfilled, Ca deficiency leads to a condition called *parturient hypocalcaemia* or milk fever (MF). MF is an economically important disease. In Tamil Nadu alone, the loss due to MF was estimated at ₹ 40.62 crores during 2005-08 (Thirunavukkarasu *et al*. 2010). Emphasis on MF is particularly important as it occurs at the most productive period of lactating dairy animals every year. It is a recurring phenomenon prominently seen in high yielding dairy cows and buffaloes. With the aim of increasing the productivity of dairy animals in India, MF incidence is bound to increase annually. MF is a gateway for many other diseases and disorders (Buragohain and Kalita 2016). With MF comes the increased risk of mastitis, retained placenta, dystocia, prolapsed uterus, metritis, delayed uterine involution, retained fetal membranes, displaced abomasum, reduced feed intake, poor rumen and intestine motility (Goff 2008; Oetzel and Miller 2012; Reinhardt *et al*. 2011).

[1] Division of Dairy Economics, Statistics and Management, ICAR–National Dairy Research Institute (ICAR-NDRI), Karnal – 132001 (Haryana), India.
* Correspondence: adeeth07@gmail.com
[2] Division of Dairy Extension, ICAR–NDRI – 132001 (Haryana), India
[3] Division of Animal Nutrition, ICAR–NDRI – 132001 (Haryana), India
[4] ICAR-Indian Institute of Wheat and Barley Research, Karnal - 132 001 (Haryana), India

Despite having a huge economic importance, evidence on the incidence of MF is poorly recorded in India. Only two studies reported the evidence from field surveys. MF incidence was reported at 12-13% in north eastern states of India (Paul *et al*. 2013) and 13-14% in Tamil Nadu (Thirunavukkarasu *et al*. 2010). We therefore complement the existing literature by reporting the MF incidence as well as estimating the milk production and total economic losses caused by the MF in Haryana. Economic loss estimation is pertinent to understand the problem in-depth and to design preventive measures accordingly. In addition, the efficiency gain by the prevention of MF is estimated.

**Materials and Methods**

We report here the baseline results from our randomized controlled trial (RCT) which aims to evaluate the impact of anionic mineral mixture supplementation on milk production and the MF[5]. Data used in this study was collected from primary and secondary sources.

*Primary data*

Survey data was collected using a pre-tested interview schedule from five villages namely Garhi Gujran (29.8370° N, 77.0172° E), Kamalpur Rodan (29.8009° N, 77.0507° E), Churni (29.8061° N, 77.0241° E), Samora (29.7327° N, 77.0452° E) and Nagla Rodan (29.8125° N, 77.0776° E) of Karnal district, Haryana during September-November 2020. Since milk fever is more prevalent in high yielding bovines and as Haryana is home for around 6.5 million female bovines with an 76% increase in exotic/crossbred cattle population (which are high yielding) during the inter census period of 2007-12 (GoI, 2012), it becomes an ideal study area to estimate the incidence of milk fever and consequent losses from it.

*Sample size estimation*

The formula to estimate the minimum detectable effect size is given below.

$$\text{Effect size} = (t_{(1-k)} + t_\alpha)\sqrt{\frac{1}{P(1-P)}} * \sqrt{\left(\frac{\sigma^2}{N}\right)}$$

This formula was used to calculate the total sample size (N) required for the study. $t_{(1-k)}$ is the power, $t_\alpha$ is the significance level, P is the proportion of treatment in the sample, $\sigma^2$ is the variance and N is the sample size (Duflo *et al*. 2007). The sample size was arrived at using the 'Optimal Design' software. A total of 200 animals from 200 dairy farmers; 100 buffaloes and 100 cows were selected from the above mentioned villages of Karnal district.

Bovines which fulfilled the following criteria were selected for the study.

1. Pregnant bovine above 2nd lactation
2. The farmers should not use any form of anionic diets[6]
3. Milk yield should be more than 10 L/day

---

[5] For details of sampling and study design, see the pre-plan registered (RCT ID - AEARCTR-0005108) in the American Economic Association's RCT registry (https://www.socialscienceregistry.org/trials/5108)
[6] The survey was part of a field experiment which aimed at evaluating the impact of anionic mineral mixture supplementation on milk fever and other outcomes

Information on socio-economic characteristics (such as age, education, family size, primary and secondary source of income, *etc*.), milk fever (awareness, precautions, incidence, death, milk loss, medicine and veterinarian's fee *etc*. in the previous lactation), insurance, herd size, disease history, veterinary services and anionic mineral mixture awareness were collected.

*Secondary data*

Information on Haryana's bovine population and proportion of animals' in-milk were collected from the 'Provisional Key Results of 20[th] Livestock Census'[7] and data on milk production was collected from the 'Basic Animal Husbandry Statistics 2019'[8].

*Estimation of Milk production losses*

$$Y_{loss} = A_{IM}.P_{MF}.Y_L.P_D[1 + S.P_{MFD}.P_{MYR}] \qquad \ldots (1)$$

Where, $Y_{loss}$ is the yearly milk lost due to MF (in L)

$A_{IM}$ is the number of animals in-milk

$P_{MF}$ is the proportion of animals affected by MF

$Y_L$ is the average lactation milk production (in L)

$P_D$ is the case fatality ratio (proportion of animals died among MF affected animals)

S is the case survival ratio $\left[S = \frac{1}{P_D} - 1 \; or \; \frac{number\ of\ animals\ survived}{number\ of\ animals\ died}\right]$

$P_{MFD}$ is the proportion of lactation days affected in milk fever affected animals

$P_{MYR}$ is the proportion of milk yield reduced per day in MF affected animals

*Estimation of Total economic losses (TEL)*

$$\text{TEL (in ₹)} = M_L + Y_V + T_C \qquad \ldots (2)$$

i.  Losses from mortality ($M_L$) (in ₹) = $[A_{IM}.P_{MF}.P_D].V$      … (3)
    Where, V is the market value of the animal (in ₹)
ii. Value of milk lost ($Y_V$) (in ₹) = $[A_{IM}.P_{MF}.Y_L.P_D[1 + S.P_{MFD}.P_{MYR}]].P$   … (4)
    Where, P is the price of milk (in ₹/L)
iii. Treatment cost ($T_C$) (in ₹) = $[A_{IM}.P_{MF}.(1 - P_D)]$ [veterinarians fee + medicine costs]      … (5)

---

[7] https://dahd.nic.in/sites/default/filess/Key%20Results%2BAnnexure%2018.10.2019.pdf
[8] https://dadf.gov.in/sites/default/filess/BAHS%20(Basic%20Animal%20Husbandry%20Statistics-2019).pdf

*Efficiency gains using economic surplus model*

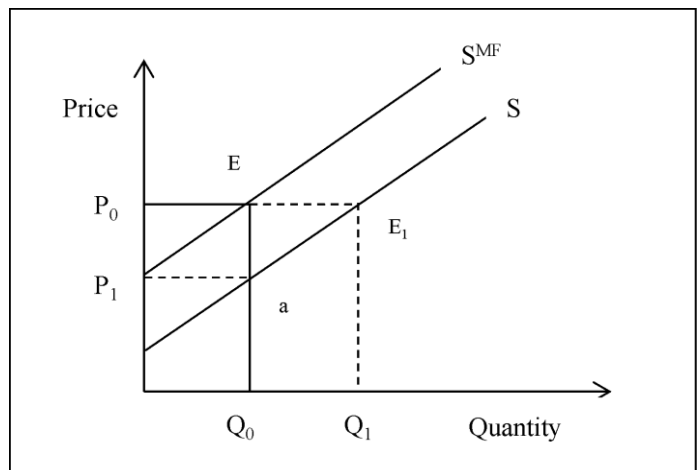

**Figure 1 Economic surplus in an open economy model with no demand restriction**

Note: In Figure 1, $S^{MF}$ represents the milk supply curve with MF (at the present level of production), S represents the new supply level if MF is prevented (increased milk production), $P_0$ is the initial price level of milk, $P_1$ is the new price level, $Q_0$ is the initial equilibrium quantity and $Q_1$ represents the quantity supplied if the MF is prevented. Area $P_0E_1aP_1$ represents the producer surplus (efficiency gain) due MF prevention.

Economic surplus model has been widely used to estimate economic losses due to diseases in animals (Alves *et al*. 2015; Bardhan *et al*. 2017; Barratt *et al*. 2018; Lhermie *et al*. 2018; Losinger 2005a, 2005b, 2006; Ott *et al*. 1995; Weldegebriel *et al*. 2009). We used it to estimate the efficiency gains to producers which might accrue if the MF is prevented. This study deviates from the standard economic surplus model which assumes a closed economy. We assume an open economy model in which excess production can be exported and the deficit could be imported. It is a realistic assumption as Haryana imports milk from the neighboring states in times of shortage. Therefore, in this setup there will be no demand restriction and the total economic surplus is equal to the producer surplus (Alston *et al*. 1995). Producer surplus can be expressed as,

$$\Delta TS = \Delta PS = K.P_0.Q_0.(1 + 0.5.Z.e) \qquad \ldots (6)$$

Where $\Delta TS$ is the change in total surplus, $\Delta PS$ the change in producer surplus, K is is the supply shift relative to the initial price ($P_0$), $Q_0$ is the initial supply (production), Z is the relative reduction in price and e is the supply elasticity of milk. K and Z is calculated as follows

$$K = \frac{e}{\% \Delta \ in \ quantity} \qquad \ldots (7)$$

$$\% \Delta \ in \ quantity = \frac{Q_1 - Q_0}{Q_0}$$

$$Z = \frac{K.e}{e + \eta} \qquad \ldots (8)$$

Where, $Q_1$ is the supply of milk if MF is completely prevented. It is calculated by adding the quantity of milk lost to the initial level of milk production by animals in-milk.

$$Q_1 = Q_0 + Y_{loss} \qquad \ldots (9)$$

$\eta$ is the absolute value of the demand elasticity (-1.035) of milk and it is taken from (Kumar *et al*. 2011). Similarly, the value of supply elasticity (0.019) was sourced from (Birthal *et al*. 2019). In a randomized evaluation of 400 dairy cattle, probability of success of anionic diets (which prevents MF) was 100%, i.e., no treated animals had hypocalcaemia (MF) (Melendez *et al*., 2019). We estimate efficiency gains at 90% success rate and 10% is excluded due to the possible mismanagement which could arise due to farmer, animal and technology related inefficiencies.

**Results and Discussion**

*Sample description*

Summary statistics of some key variables of milk production in the sample are presented in Table 1. All the bovines were above second lactation as we had only selected animals of higher order lactation. Peak milk yield in buffaloes was around 12 L/d and 16-17 L/d in cows with average herd size of 5-6 animals. Average green fodder, dry fodder and concentrate intake was 19.94, 11.82 and 3.70 kg/d respectively with buffaloes consuming a little more than cows. Mineral mixture feeding was similar in both cows and buffaloes (0.03 kg/d). Area under fodder was slightly higher among buffalo than cow owners while labour used was a tad higher for cows than buffaloes.

**Table 1 Description of milk production in the sample**

| Variables | Buffalo (n=105) | | Cow (n=107) | | Combined (N=212) | |
|---|---|---|---|---|---|---|
| | **Mean** | **S.E.** | **Mean** | **S.E.** | **Mean** | **S.E.** |
| Parity (nos.) | 2.74 | 0.06 | 2.81 | 0.08 | 2.78 | 0.05 |
| Peak yield in previous lactation (L/animal/d) | 12.06 | 0.23 | 16.01 | 0.56 | 14.06 | 0.33 |
| Peak yield in present lactation (L/animal/d) | 12.38 | 0.33 | 16.67 | 0.85 | 14.69 | 0.52 |
| Herd size (nos.) | 6.30 | 0.29 | 4.90 | 0.27 | 5.59 | 0.20 |
| Green fodder (kg/animal/d) | 20.76 | 0.63 | 19.13 | 0.53 | 19.94 | 0.42 |
| Dry fodder (kg/animal/d) | 12.20 | 0.46 | 11.44 | 0.45 | 11.82 | 0.32 |
| Concentrates (kg/animal/d) | 3.82 | 0.15 | 3.57 | 0.15 | 3.70 | 0.11 |
| Mineral mixture (kg/animal/d) | 0.03 | 0.00 | 0.03 | 0.00 | 0.03 | 0.00 |
| Fodder area (acres) | 0.62 | 0.06 | 0.60 | 0.05 | 0.61 | 0.04 |
| Labor (nos.) | 2.25 | 0.07 | 2.42 | 0.10 | 2.33 | 0.06 |

*Milk fever*

Responses elicited from the dairy farmers regarding MF are summarized in Figure 2. A 3/4[th] of the dairy farmers were aware of MF while only 44% (in buffaloes) and 55% (in cows) of them were taking precautionary measures. Precautionary measures (according to the dairy farmers) include feeding jaggery (unrefined cane sugar)[9], calcium supplements and mineral mixtures. Incidence of clinical MF in the sample area was higher among high yielding cows (28%) than buffaloes (19%) while mortality rate was similar at around 2%. The MF incidence numbers seems higher than previous literature (Paul *et al*. 2013; Reinhardt *et al*. 2011; Thirunavukkarasu *et al*. 2010) because of the sample selection; we have selected only high yielding animals above second parity. The mortality rate was lower because of the ease of treating MF affected animals; calcium injections or intravenous (IV) infusions.

---
[9] https://en.wikipedia.org/wiki/Jaggery

Predictive margins were estimated (using a logit model) to assess the differential incidence of MF by parity and type of animals. The results are presented in Appendix 1 and Figure 3. It was found that the probability of MF incidence was higher among the high yielding cows than buffaloes at all the parities. The highest probability of MF incidence was found at 4$^{th}$ and 5$^{th}$ parities of cows. MF incidence had a significant positive relationship with the parity. In other words, as the parity increases, probability of MF incidence increases. Therefore, the older, high yielding animals should be given extra care to prevent MF from occurring.

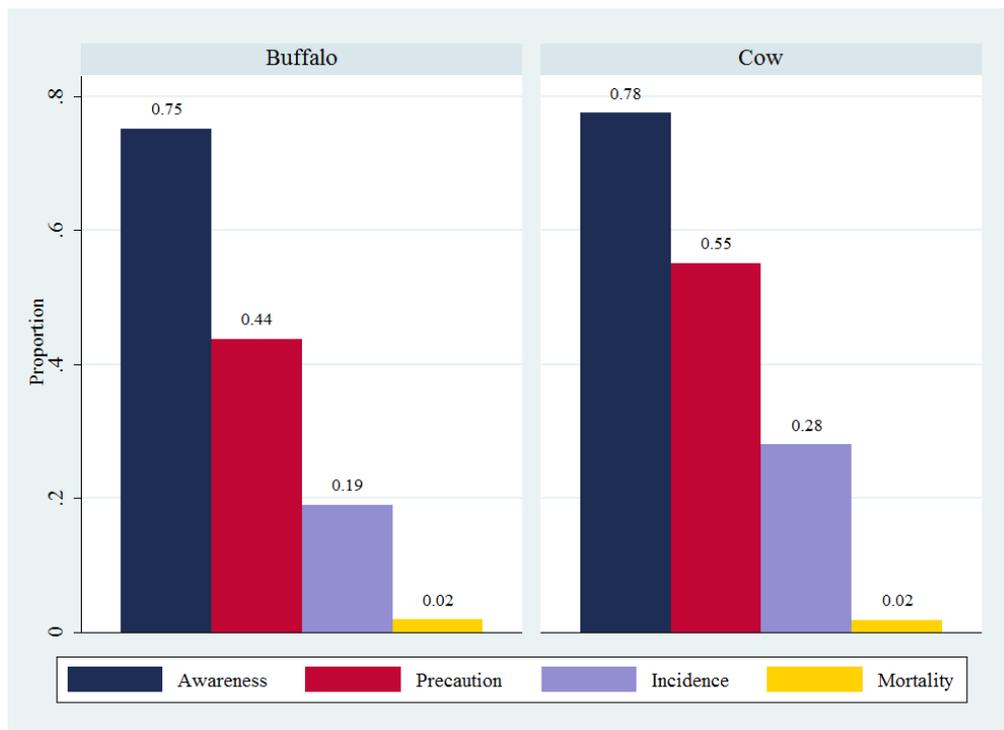

**Figure 2 Milk fever: awareness, precaution, morbidity and mortality in sample bovines**

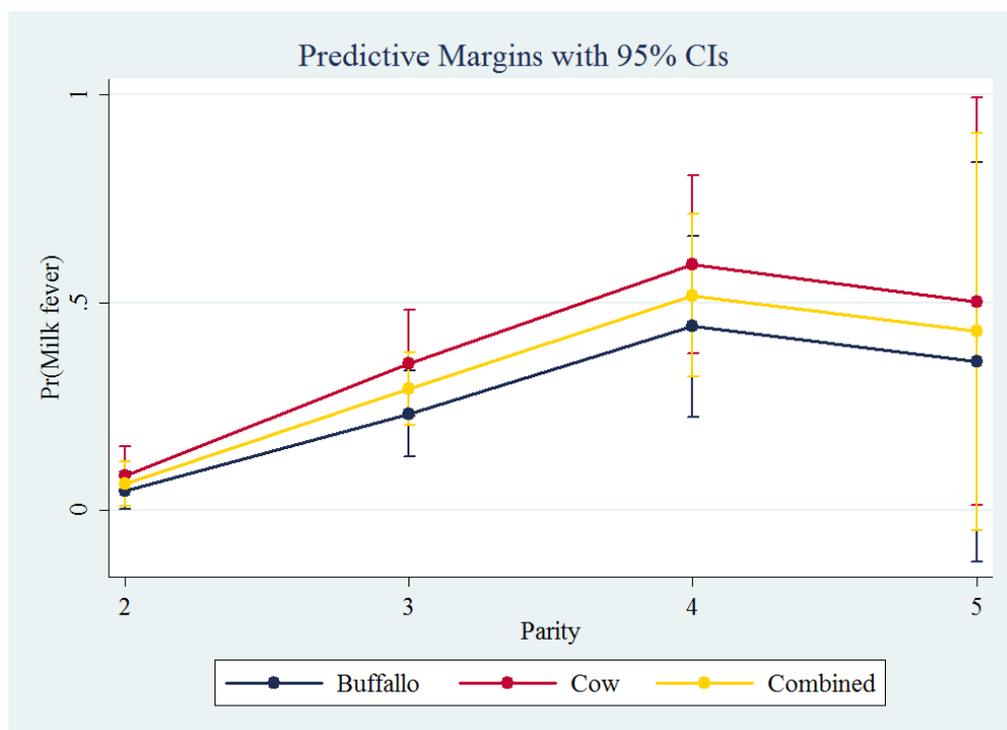

**Figure 3 Parity-wise predictive margins of milk fever incidence in sample bovines**

*Milk production and total economic loss*

Panel (a) in Table 2 summarizes the means of variables used and panel (b) shows the estimated milk production and economic losses due to MF. In the sample, quantity of milk production losses was 7.72 tonnes among cows and 6.15 tonnes among buffaloes. Total economic losses were however higher in buffaloes (₹ 4.63 lakhs, *i.e.*, ₹ 3915/animal) than cows (₹ 4.18 lakhs *i.e.*, ₹ 4413/animal); because of higher milk prices and market value of buffaloes.

Assuming that the MF incidence and mortality due to MF will be similar in the sample area and Haryana, the milk production and economic losses were also estimated for Haryana state of India. Milk production and economic losses were highest among buffaloes in Haryana. Total economic losses due to MF in Haryana were estimated at ₹ 999.9 crores with ₹ 914.2 crores of losses in buffaloes alone.

Value of milk lost had the highest share of around 60% in total economic losses followed by losses from mortality of animals (29%) and treatment costs (12%) in both the sample and Haryana. Further, the values and shares of milk and mortality losses were highest in case of buffaloes because of the higher milk prices and market value of buffaloes.

The economic losses to the tune of ₹ 1000 crores can be reduced by just taking preventive measures. MF is not a disease, it does not spread, and it is a nutritional disorder which could be prevented with relative ease. For example, there is overwhelming evidence that pre-partum anionic diets prevent MF (Aasif Shahzad *et al.* 2008; Charbonneau *et al.* 2006; Lean *et al.* 2019; Melendez *et al.* 2019; Wu *et al.* 2008). Such preventive measures could be taken in India to prevent huge economic losses due to MF. For instance, anionic diets available in India[10] is sold at ₹ 180/kg[11]. Feeding this to pregnant bovines a month prior to parturition costs around ₹ 540 per animal (@ 100g for 30 days – 3 kg for an animal). The treatment cost per animal (₹ 2500 see Table 2) is around five times higher than the prevention cost (₹ 540). Total cost for preventing MF (if fed to all animals in-milk) is around ₹ 127 crores. Therefore, economic losses due to MF could be prevented by spending a mere 1/8$^{th}$ of the amount of total economic losses.

In the next section we present what happens if the milk production losses are prevented in the advent of a technology. We estimate the economic efficiency at different levels of adoption of a technology which could prevent MF.

**Table 2 Estimated milk production and economic losses due to MF in sample villages and Haryana**

| | Description | Sample (n=212) | | | Haryana (N=3250862) | | |
|---|---|---|---|---|---|---|---|
| | | Cows | Buffaloes | Total | Cows | Buffaloes | Total |
| a. | **Means of data used for loss estimation** | | | | | | |
| T | Animal (nos.) | 107.00 | 105.00 | 212.00 | 485603.00 | 2765259.00 | 3250862.00 |
| $P_{IM}$ | Proportion of animals in-milk | 1.00 | 1.00 | 1.00 | 0.73 | 0.72 | 0.73 |
| $A_{IM}$ | In-milk animal population (nos.) | 107.00 | 105.00 | 212.00 | 354490.19 | 1990986.48 | 2345476.67 |
| $P_{MF}$ | Proportion of MF affected animals | 0.28 | 0.19 | 0.24 | 0.28 | 0.19 | 0.24 |
| $P_{DMF}$ | Proportion of animals died | 0.02 | 0.02 | 0.02 | 0.02 | 0.02 | 0.02 |
| $P_D$ | Case fatality ratio | 0.07 | 0.10 | 0.08 | 0.07 | 0.10 | 0.08 |
| D | Death (nos.) | 2.00 | 2.00 | 4.00 | 6629 | 38028 | 44657 |
| A | Morbidity (nos.) | 30.00 | 20.00 | 50.00 | 99399.05 | 379282.92 | 478681.97 |
| Y | Average daily milk yield per milch animal (L/d) | 10.06 | 8.56 | 9.31 | 8.92 | 9.11 | 9.02 |
| $Y_L$ | Average lactation milk production (L) | 3068.30 | 2610.80 | 5679.10 | 2720.60 | 2778.55 | 5499.15 |
| S | Case survival ratio | 14.00 | 9.00 | 11.50 | 13.99 | 8.97 | 11.48 |
| $P_{MFD}$ | Proportion of lactation days affected | 0.02 | 0.02 | 0.02 | 0.02 | 0.02 | 0.02 |
| $P_{MYR}$ | Proportion of milk lost during affected days | 0.80 | 0.86 | 0.83 | 0.80 | 0.86 | 0.83 |
| P | Price of milk (₹/L) | 30.00 | 45.00 | 37.50 | 30.00 | 45.00 | 37.50 |
| V | Market value of animal (₹) | 53333.00 | 74250.00 | 63791.50 | 53333.00 | 74250.00 | 63791.50 |
| TC | Treatment cost per animal (₹) | 2882.00 | 2115.00 | 2498.50 | 2882.00 | 2115.00 | 2498.50 |
| b. | **Estimated milk production and economic losses** | | | | | | |
| $Y_{LOSS}$ | Milk production losses (tonnes) | 7.72 | 6.15 | 14.29 | 22674.61 | 124377.44 | 154499.27 |
| $Y_V$ | Value of milk lost (₹ crores) | 0.023 (55.27) | 0.028 (59.73) | 0.054 (58.53) | 68.02 (52.28) | 559.70 (61.22) | 579.37 (57.94) |
| $T_C$ | Cost of treatment (₹ crores) | 0.008 (19.27) | 0.004 (8.22) | 0.011 (12.49) | 26.74 (20.55) | 72.17 (7.89) | 126.46 (12.65) |
| $M_L$ | Losses from mortality (₹ crores) | 0.011 (25.47) | 0.015 (32.05) | 0.027 (28.99) | 35.35 (27.17) | 282.36 (30.88) | 294.07 (29.41) |

---

[10] https://kamdhenufeeds.com/product/anionic-mishran/
[11] https://www.indiamart.com/proddetail/anionic-mishran-afs-feeds-supplement-6062198248.html

| | | | | | | | |
|---|---|---|---|---|---|---|---|
| **TEL** | Total economic losses due to MF (₹ crores) | 0.042 (100.00) | 0.046 (100.00) | 0.092 (100.00) | 130.11 (100.00) | 914.23 (100.00) | 999.91 (100.00) |

Note: Figures in parenthesis are per cent to total economic losses
Source: For sample – author's estimation based on field survey; For Haryana – population data was collected from the 20[th] livestock census (2019) and milk yield was collected from Basic Animal Husbandry Statistics 2019

*Efficiency gains when MF is prevented*

Following (Alston *et al*. 1995; Losinger 2005b, 2005a, 2006), a change in unit cost of production (due to decrease in treatment costs of prevented MF) is modeled as a parallel shift in supply curve (increased milk production) and the difference between these static situations (with MF and without MF) gives us the measure of economic efficiency generated due to prevention of MF. Efficiency gains to the producers could be thought of an increase in efficiency due to increased milk production at decreased production costs taking into account the market dynamics (demand and supply elasticities).

**Table 3 Estimated efficiency gain (producer surplus) if milk fever is prevented in Haryana**

| Variables | Cows | Buffaloes | Total |
|---|---|---|---|
| Supply elasticity (e) | 0.019 | 0.019 | 0.019 |
| Demand elasticity ($\eta$) | -1.035 | -1.035 | -1.035 |
| In-milk animals (million) | 0.354 | 1.991 | 2.345 |
| Proportion of MF affected animals | 0.280 | 0.191 | 0.235 |
| Proportion of animals died | 0.019 | 0.019 | 0.019 |
| Milk yield (L/d) | 8.920 | 9.110 | 9.015 |
| Milk Loss (000 tonnes) | 22.675 | 124.377 | 154.499 |
| Milk production (000 tonnes) | 252.390 | 948.194 | 1200.585 |
| Milk production without MF (000 tonnes) | 275.065 | 1072.572 | 1355.084 |
| % change in supply | 0.090 | 0.131 | 0.129 |
| Supply shift relative to the initial equilibrium price (K) | 4.728 | 6.904 | 6.773 |
| Relative reduction in price (Z) | 0.085 | 0.124 | 0.122 |
| Price (₹/L) | 30.000 | 45.000 | 37.500 |
| Success rate | 0.900 | 0.900 | 0.900 |
| **Efficiency gain (₹ crores)** | **3224.8** | **26543.4** | **27475.8** |

Description of all the variables used and the estimated efficiency gain is presented in Table 3. If, any technology is 90% successful in preventing MF and if the technology is adopted by all the dairy farmers in Haryana, a total of ₹ 27476 crores of economic efficiency (producer welfare) could be gained. A 100% adoption rate could be unrealistic; therefore we estimated the efficiency gains at different level of adoption rates (Figure 4). Even at a lower bound scenario of 20% adoption of MF technology, there will be a total efficiency gain of ₹ 54950 crores. Agricultural technology adoption literature suggests that there will be an average of 40% adoption in early stages, and at that rate, there will be an efficiency gain of ₹ 10,990 crores for dairy farmers in Haryana. At a higher rate of adoption (60%) the efficiency gains could be ₹ 16,485 crores. Therefore, this study highlights the fact that taking preventive measures not only reduces the losses caused by MF, but also creates a substantive amount of efficiency to the dairy farmers.

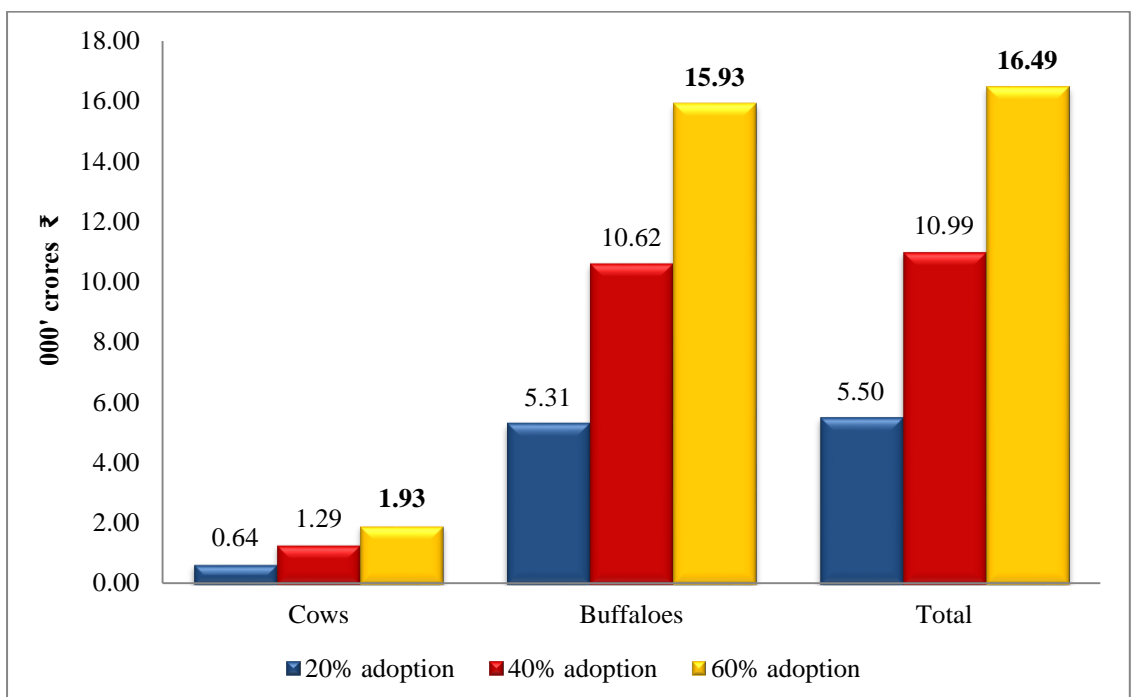

**Figure 4 Efficiency gains to milk producers at different level of adoption (at 90% success rate)**

To conclude, among 212 animals, incidence of clinical MF in the sample area was higher among high yielding cows (28.04%) than buffaloes (19.05%) while the mortality rate was similar at around 2%. Assuming similar incidence and mortality rates in the study region, the total economic losses was estimated at ₹ 999.91 crores (~US $ 137 million) in Haryana, India. Value of milk lost had the highest share of around 60% in total economic losses, followed by losses from mortality of animals (29%) and treatment costs (12%). There are a few important caveats to keep in mind. First, the incidence and mortality numbers could be over-estimates because of the selection of higher lactation animals for the study. Second is the assumption of similar incidence and mortality rates in sample and Haryana.

Given the economic importance of MF, this study estimated the efficiency gains if the MF is prevented at the advent of a technology. Using Economic Surplus model which accounts for market dynamics of milk, efficiency gain at 40% adoption rate was estimated at ₹ 10,990 crores (~US $ 1.5 billion). Therefore the study calls for taking up preventive measures against the MF. For instance, total cost of preventing MF (if all the animals' in-milk is fed anionic diets) is around ₹ 127 crores (~US$ 17 million). Economic losses due to MF could thus be prevented by spending a mere 1/8$^{th}$ of the total economic losses which could result in more efficient milk production.

**Disclaimer**
The views expressed here are of the authors and not of the affiliated institutions.

## Declarations

**Funding** ICAR-NDRI Farmers First Programme.

**Conflicts of interest/Competing interests:** Authors have no conflicts of interest/competing interests.

**Ethics approval and consent to participate**

All the dairy farmers were informed and gave their consent before conducting the interview.

**Appendix**

**Appendix 1 Predictive margins of milk fever**

| Dependent variable – milk fever (1/0) | Margin | Delta-method Std. Err. | z | P>|z| |
|---|---|---|---|---|
| Lactation order (Parity) | | | | |
| 2 | 0.06** | 0.03 | 2.32 | 0.02 |
| 3 | 0.29*** | 0.04 | 6.56 | 0.00 |
| 4 | 0.52*** | 0.10 | 5.17 | 0.00 |
| 5 | 0.43* | 0.24 | 1.76 | 0.08 |
| Animal (Buffalo=0, Animal=1) | | | | |
| 0 | 0.19*** | 0.04 | 5.03 | 0.00 |
| 1 | 0.28*** | 0.04 | 6.56 | 0.00 |
| Interaction term (Lactation order # animal) | | | | |
| 2 # 0 (2$^{nd}$ parity buffalo) | 0.05** | 0.02 | 2.01 | 0.04 |
| 2 # 1 (2$^{nd}$ parity cow) | 0.08** | 0.04 | 2.23 | 0.03 |
| 3 # 0 (3$^{rd}$ parity buffalo) | 0.23*** | 0.05 | 4.38 | 0.00 |
| 3 # 1 (3$^{rd}$ parity cow) | 0.35*** | 0.07 | 5.41 | 0.00 |
| 4 # 0 (4$^{th}$ parity buffalo) | 0.44*** | 0.11 | 3.98 | 0.00 |
| 4 # 1 (4$^{th}$ parity cow) | 0.59*** | 0.11 | 5.40 | 0.00 |
| 5 # 0 (5$^{th}$ parity buffalo) | 0.36 | 0.25 | 1.46 | 0.15 |
| 5 # 1 (5$^{th}$ parity cow) | 0.50** | 0.25 | 2.01 | 0.04 |